\documentclass[epj
]{svjour}

\usepackage{epsfig}
\usepackage[american]{babel}
\usepackage{amsfonts} 
\usepackage[dvips]{color}
\usepackage[latin2]{inputenc}
\usepackage[T1]{fontenc}
\usepackage{graphicx}
\usepackage{graphics}
\usepackage{amsmath}
\usepackage{amsfonts}
\usepackage{amssymb}

\begin{document}

\title{Bipolarons and polarons in the Holstein-Hubbard model: Analogies and differences}

\author{O. S. Bari\v si\' c\inst{1} \and S. Bari\v si\' c\inst{2}
}                     
%
%
\institute{Institute of Physics, Bijeni\v cka c. 46, HR-10000 Zagreb, Croatia\and Department of Physics, Faculty of Science, University of
Zagreb, Bijeni\v cka c. 32, HR-10000 Zagreb, Croatia}
%
%

\abstract{The single bipolaron problem is examined in the context of the 1D Holstein-Hubbard model, emphasizing analogies and differences with respect to the complementary single polaron physics. The bipolaron band structure below the phonon threshold is revealed, showing a complex relationship between numerous excited bands as the adiabatic limit is approached. Light bipolarons with significant binding energy, the stability of large bipolarons, the small to large bipolaron crossover as a function of the Hubbard repulsion, as well as the bipolaron dissociation, are investigated in detail, disentangling adiabatic, nonadiabatic and lattice coarsening effects. It is emphasized that condensation of bipolarons occurs in the dilute limit only at very low temperatures.
} 
\maketitle

\maketitle

\section{Introduction}

The polaron represents a quasi-particle involving a single electron coupled to the lattice degrees of freedom. This coupling introduces local correlations between the electron and the lattice field, characterized by a finite electron-phonon correlation length $d_{pol}$ and an effective mass $m_{pol}$. The translational symmetry of the lattice is preserved because the electron and the lattice field (deformation) can travel together. When the local Coulomb repulsion is not too strong, a pair of electrons can bind by sharing a common lattice deformation field. Such an entity is called a bipolaron. The bipolaron condensation energy $\Delta_{bp}$ is defined as the difference in binding energies of the bipolaron $E_{bp}$ and two uncorrelated polarons $E_{pol}$, $\Delta_{bp}=2E_{pol}-E_{bp}$. The bipolaron mass $m_{bp}$ is associated with a joint motion along the lattice of a pair of electrons and the corresponding lattice field.

The present work is focused on low-frequency properties of a single bipolaron in the context of the Holstein-Hubbard (HH) model \cite{Anderson}. Previous investigations of bipolarons based on this model range from variational studies \cite{Magna,Proville,Proville2,Filippis,Eagles,Sil} to various numerical approaches, such as exact diagonalization calculations for finite clusters \cite{Marsiglio,Mello,Crljen,Wellein} and the infinite lattice \cite{Bonca,Shawish}, quantum Monte-Carlo simulations \cite{Raedt,Macridin,Hohenadler2}, the density matrix renormalization group \cite{Zhang}, and the cluster perturbation theory \cite{Hohenadler}. Generalizations of the Holstein-Hubbard model have been studied \cite{Bonca002,Hague} by varying the range of the electron-phonon interaction. Perturbative expansion around the atomic limit is analyzed comprehensively in reference~\cite{Bonca010}, whereas the exact Green's function for the two-sites problem is derived analytically in reference~\cite{Berciu}. A broader overview of the bipolaron related physics may be obtained from references~\cite{Emin003}, \cite{Mott} and \cite{Devreese}.

Although a number of methods have been developed in the last decade to investigate the HH bipolaron problem, the description of an important part of the phase diagram, which involves large adiabatic bipolarons and associated crossovers to other regimes, is still scant. The problem lies in electron-phonon correlations that emerge when the coupling is strong and the electron-phonon correlation length $d_{bp}$ is large, $d_{bp}/a\gg1$ ($a$ is the lattice constant). Namely, such correlations impose particular demands on an accurate, quantum treatment of bipolarons since the correlations with many phonons at large distances from two electrons have to be taken into account. 

For exact diagonalization approaches \cite{Bonca,Shawish,Barisic10} that are based on a truncation of the Hilbert space, the number of states that should be considered grows exponentially with increasing coupling and increasing $d_{bp}$. This limits the accessibility of the phase diagram for such calculations. Indeed, to best of our knowledge, all present numerically accurate studies of the quantum HH bipolarons are restricted either to weak couplings or to limited values of the adiabatic ratio $t/\omega_0\lesssim2$, where $t$ is the electron hopping energy and $\omega_0$ is the optical phonon energy of the HH model. In other words, the formation and properties of large bipolarons with significant binding energy still represents a challenging subject.

For the half-filling case, the bipolaronic phase has been studied by the dynamical mean field theory (DMFT) \cite{Capone}. However, this particular case of high charge concentration necessarily differs from the dilute limit, when the single-bipolaron theory applies. Furthermore, since its diagrammatic expansion neglects vertex corrections involving phonons at different lattice sites, the DMFT for finite-dimensional systems fails to describe properly the adiabatic electron-lattice correlations spanning several lattice sites \cite{Barisic11}. For the HH model, with local coupling and local phonons, this problem is most pronounced for the one-dimensional $D=1$ system because, in this case, large adiabatic (bi)polarons are stable, while such correlations for higher dimensional $D>1$ systems \cite{Kabanov,Kalosakas} are short-lived. 

The treatment developed here combines the results found in the adiabatic limit with numerical results obtained by the recently proposed relevant coherent state method (RCSM) \cite{Barisic5}. Within the RCSM, (bi)polaron states are obtained by solving a generalized eigenstate problem. The latter is defined by choosing a trial set of the most relevant wave functions after a careful analysis of the properties of the low-frequency adiabatic and nonadiabatic correlations.

The RCSM offers an improvement over previous calculations in several different ways. First, it can be applied to any value of the adiabatic ratio $t/\omega_0$, including the $t/\omega_0\gg1$ part of the bipolaron phase diagram where the strongly-coupled large bipolarons emerge. Second, the method provides the full low-frequency band structure of the bipolaron states including the excited coherent bands, which, to the best of our knowledge, have not been previously reported. Such analyses allow a detailed description of the bipolaron low-frequency dynamics and extend the previous investigations, which were mainly focused on the properties of the ground state, to include the effective mass and, in some cases, the dispersion of the lowest bipolaron band. The existence of the first excited zero-momentum state below the phonon threshold for inelastic scattering had been demonstrated in reference~\cite{Bonca002}. 

The importance of a successful treatment of the long-range adiabatic correlations can easily be set in a broader context, involving models for which the electron-phonon coupling is not purely local (on-site) as in the HH model. Namely, for any dimension $D$, the increasing range of the electron-phonon interaction results in an increment of the correlation length $d_{bp}$. For $D>1$, this increment necessarily introduces large adiabatic bipolarons into the phase diagram. In this respect the RCSM, used here for $D=1$, has additional advantages since it can be implemented with a high accuracy to a broad range of models and $D>1$.

The present paper is organized as follows. After a short introduction to the problem in Section~\ref{Sec0}, Section~\ref{Sec2} covers the single bipolaron case, starting with the adiabatic limit and emphasizing analogies with adiabatic polarons. In the next step, differences between the polarons and bipolarons appearing with nonadiabatic correlations are examined. This is followed by an analysis of the role of the Hubbard repulsion $U$ on the condensation energy for both large and small adiabatic bipolarons. In Section~\ref{Sec3} the RCSM  is used to calculate accurately the bipolaron spectra below the phonon threshold for the incoherent scattering. Beside the lowest coherent bipolaron band, additional coherent excited bands are found to be related to excitations of adiabatically softened phonon modes of the moving lattice field. Particular attention is paid to the small and large $U$ adiabatic regimes and the parameter space in which small light bipolarons are formed with substantial condensation energies. Section~\ref{Sec1} gives a brief discussion of finite temperature effects, with emphasis on particular aspects of the dilute limit. A summary of the results is given in Section~\ref{SecSummary}.

\section{General\label{Sec0}}

The 1D Holstein-Hubbard Hamiltonian is given by \cite{Anderson}

\begin{eqnarray}
\hat{H}&=&\hat{T}+\omega_0\sum_{n}u^2_n-t\sum_{n,s}c_{n,s}^{\dagger}\;
(c_{n+1,s}+c_{n-1,s})
\nonumber\\&-&
2g\sum_{n,s} c_{n,s}^{\dagger}c_{n,s}\;u_n
+U\sum_{n,s\neq s'}\hat c_{n,s}^{\dagger}c_{n,s} c_{n,s'}^{\dagger}c_{n,s'}
\label{HolHubHam}\;,
\end{eqnarray}

\noindent where $c_{n,s}$ is the annihilation electron operator on site $n$ with spin $s=\uparrow,\downarrow$, $u_n$ is the dimensionless lattice displacement for the site $n$, and $\hat T$ represents the lattice kinetic energy shifted, for convenience, by the zero-point energy of the free lattice,

\begin{equation}
\hat{T}=-\omega_0\sum_n \left(\frac{1}{4}\frac{\partial^2}{\partial u^2_n}+\frac{1}{2}\right)\;.\label{Tlatt}
\end{equation}

\noindent The Hamiltonian (\ref{HolHubHam}) describes electrons in the tight binding nearest-neighbor approximation coupled to the dispersionless branch of optical phonons. The electron-phonon and electron-electron interactions, given by the last two terms in equation~(\ref{HolHubHam}), are local. While the latter interaction is instantaneous, the former contains retardation effects through the lattice coordinates $u_n$. $u_n=1/2$ corresponds to the zero-point displacement of free lattice oscillators. We use $\omega_0$ as the unit of energy.

In the same way as for polarons \cite{Barisic8}, two fundamentally different kinds of correlative behavior between the electron and lattice subsystems can be distinguished for HH bipolarons. Adiabatic correlations are described by an electron pair that instantaneously adjusts to the motion of the lattice deformation field. On the other hand, during nonadiabatic processes, the electrons temporarily detach from the lattice field. 

It is frequently assumed for $U=0$ that the parameter $t/\omega_0$ is sufficient to distinguish between the regime dominated by the nonadiabatic correlations from the one dominated by the adiabatic correlations. However, it is important to stress that this distinction should also account for the strength of the electron-phonon coupling $g/\omega_0$. That is, the adiabatic correlations develop only for sufficiently strong couplings, whereas the weak-coupling limit, irrespectively of $t/\omega_0$, involves purely nonadiabatic dynamics \cite{Barisic8}. For a fixed $t/\omega_0$, the latter can always be reached by decreasing $g/\omega_0$.

One important ingredient of the bipolaron physics are lattice coarsening (discreteness) effects. The role of these effects is determined by the electron-phonon correlation length $d_{bp}$. By analogy with the polaron case \cite{Barisic8}, the discreteness of the lattice deformation pins small bipolarons ($d_{bp}\approx a$), whereas in the opposite limit of large bipolarons ($d_{bp}\gg a$), the continuum approximation can be invoked. 

\section{Adiabatic approximation\label{Sec2}}

In the adiabatic limit the electron part of the bipolaron wave function behaves as if it commutes with the lattice kinetic energy (\ref{Tlatt}), depending parametrically on time, through the lattice deformation,

\begin{equation}
|\eta(\vec u)\rangle=\sum_{n,m}\eta_{n,m}\;
c_{n,\uparrow }^\dagger c_{m,\downarrow}^\dagger|0\rangle\;,\label{VFBiPol}
\end{equation}

\noindent where, for the sake of brevity, the lattice deformation is denoted by an $N$-dimensional vector $\vec u\equiv\{u_n\}$, with $N$ being the number of lattice sites ($N\rightarrow\infty$). 

For the singlet spin configuration, the two-electron wave function (\ref{VFBiPol}) satisfies $\eta_{n,m}=\eta_{m,n}$, whereas for the triplet it satisfies $\eta_{n,m}=-\eta_{m,n}$. Furthermore, in the $U\rightarrow\infty$ limit the singlet solution becomes degenerate with the triplet solution. Since in the current, as well as in other works \cite{Bonca,Bonca002} no indication of stable $U=\infty$ singlet bipolarons is found, it is expected that the triplet bipolarons are unstable in the whole parameter space of the 1D HH model.

With appropriate normalization $\sum_{n,m}\eta_{n,m}^*\eta_{n,m}=1$, the expectation value of the Hamiltonian (\ref{HolHubHam}) with respect to $|\eta(\vec u)\rangle$ is obtained as

\begin{equation}
\hat H_{AD}=\hat T+\omega_0\;\vec u^2+\varepsilon_{AD}(\vec u)\;,\label{BiLocE}
\end{equation}

\noindent where the adiabatic electron energy $\varepsilon_{AD}(\vec u)$ is given by the ground-state ($i=0$) solution $\varepsilon^{(0)}(\vec u)$ of 

\begin{eqnarray}
&&\varepsilon^{(i)}(\vec u)\;\eta_{n,m}=-2g\;(u_n+u_m)\;\eta_{n,m}+U\;\delta_{n,m}\;\eta_{n,n}\nonumber\\
&&\;\;-t\;
(\eta_{n+1,m}+\eta_{n-1,m}+\eta_{n,m+1}+\eta_{n,m-1})\;.
\label{NLSEBiPol}
\end{eqnarray}

\noindent The last two terms in equation~(\ref{BiLocE}) are functions of $\vec u$. This means that they can be interpreted as the lattice potential energy, henceforth referred to as the adiabatic potential $U_{AD}(\vec u)$. $\varepsilon_{AD}(\vec u)$ defines the change of the lattice potential energy due to the adiabatic correlations, with respect to the free-lattice case. The excited-state solutions ($i>0$) of equation~(\ref{NLSEBiPol}) may be used to analyze nonadiabatic effects.

\subsection{Adiabatic Holstein ($U=0$) bipolaron vs. polaron}

For $U=0$, the electron part of the adiabatic wave function can be factorized as a product of single-electron wave functions. Thus, the energy of the electron subsystem is given by the sum of single-electron energies $\varepsilon_{el}^{(i)}$, the latter being solutions of 

\begin{equation}
\varepsilon_{el}^{(i)}\;\eta_n=-t\left(\eta_{n+1}+\eta_{n-1}\right)
-2g\; u_n\;\eta_n\;,
\label{Lam}
\end{equation}

\noindent with $\eta_n$ the single-electron wave function.

At this point, it is convenient to make use \cite{Barisic8} of the sum rule

\begin{equation}
u_n=N_{el}\;\frac{g}{\omega_0}\tilde u_n \;,\;\;\;\sum_n\tilde u_n=1\;,\label{sumrule} 
\end{equation}

\noindent where $N_{el}$ is the number of electrons in the system.
This sum rule follows from the fact that the homogenous $q=0$ lattice mode couples only to the total electron density \cite{Feinberg}, which is fixed. Introducing $\tilde u_n$ and $\Lambda^{(i)}=\varepsilon_{el}^{(i)}/N_{el}\varepsilon_p$ in equation~(\ref{Lam}) as rescaled quantities, 

\begin{equation}
\Lambda^{(i)}\;\eta_n=-\frac{1}{N_{el}\;\lambda}\left(\eta_{n+1}+\eta_{n-1}\right)
-2\;\tilde u_n\;\eta_n\;,
\label{Lamtilde}
\end{equation}

\noindent with parameters $\varepsilon_p$ and $\lambda$ given by $\varepsilon_p=g^2/\omega_0$ and $\lambda=\varepsilon_p/t$, the adiabatic potential $U_{AD}(\vec u)$ rewritten in terms of those quantities takes the form 

\begin{equation}
U_{AD}(\vec u)= \varepsilon_p\;\left(N_{el}^2\sum_n \tilde u_n^2+N_{el}\sum_{i=0}^{N_{el}}\Lambda^{(i)}(\tilde u_n, N_{el}\;\lambda)\right)\;.\label{UADU0}
\end{equation}

\noindent Here, the summation over $i$ involves the lowest $N_{el}$ occupied single-electron states of equation~(\ref{Lamtilde}).

Equation (\ref{UADU0}) is general and valid for any $N_{el}$. In particular, for the bipolaron ($N_{el}=2$) and the polaron ($N_{el}=1$) case, only the ground ($i=0$) state of the electron spectrum (\ref{Lamtilde}) contributes to $U_{AD}(\vec u)$. It is singly occupied for the polaron and, due to the spin degeneracy, doubly so for the bipolaron. An important consequence of this property is that $U_{AD}(\vec u)$ for $N_{el}=2$ exhibits the same behavior as for $N_{el}=1$ with four times larger $\varepsilon_p$ and twice as large $\lambda$,

\begin{equation}
\varepsilon_p\leftrightarrow 
4\varepsilon_p\;,\;\;\;\lambda\leftrightarrow
2\lambda\;,\;\;\;g\leftrightarrow2g\;,\;\;\;t\leftrightarrow 
2t\;.\label{substi}
\end{equation}

\noindent Thus, the adiabatic Holstein bipolaron problem can be simply mapped through (\ref{substi}) to the adiabatic Holstein polaron problem. This useful property has not been noted previously.

The quantity in the brackets on the right hand side of equation~(\ref{UADU0}) depends only on one parameter, $\lambda$. As known from the polaron theory \cite{BBarisic}, $\lambda$ defines the adiabatic electron-phonon correlation length (the polaron size), $d_{pol}/a\approx1+2/\lambda$. Using equation~(\ref{substi}), one finds that for the same parameters the adiabatic Holstein bipolaron is always smaller than the polaron, i.e., $d_{bp}/a\approx1+1/\lambda$. This also means that the pinning effects due to the discreteness of the lattice field are stronger for bipolarons than for polarons. In addition, the bipolarons are characterized by a doubled lattice deformation in equation~(\ref{sumrule}), which renders them heavier than polarons.

\subsection{Nonadiabatic corrections for $U=0$}

While the simple mapping (\ref{substi}) exists between adiabatic bipolarons and polarons, such an analogy is absent for nonadiabatic corrections involving the excited states of the adiabatic electron spectrum (\ref{NLSEBiPol}). For polarons, the ground ($i=0$) and excited states ($i>0$) of the adiabatic electron spectrum are given by

\begin{equation}
\varepsilon_p\;\Lambda^{(i)}(\tilde u_n, \lambda)\;,\label{elpolsp}
\end{equation}

\noindent whereas the bipolaron case involves one- and two-electron excitations,
 
\begin{equation}
2\varepsilon_p\;\left[\Lambda^{(i)}(\tilde u_n, 2\lambda)+\Lambda^{(j)}(\tilde u_n, 2\lambda)\right]\;. \label{elbisp}
\end{equation}

The comparison of the two spectra in equations~(\ref{elpolsp}) and (\ref{elbisp}) reveals an important physical property: for the same parameters the bipolaron is "more adiabatic" than the polaron, because the former is characterized by a larger gap $\Delta_\eta$ in the electron spectrum between the ground and excited states. In this respect, it is instructive to consider two opposite limits, the large and the small (bi)polaron limit, corresponding to $\lambda\ll1$ and $\lambda\gg1$, respectively. For large adiabatic polarons, $d_{bp}\gg a$, the gap in the electron spectrum $\Delta_\eta$ can be evaluated in the continuum approximation \cite{Turkevich,Barisic8}, which gives $\Delta_\eta^{pol}=\varepsilon_p\;\lambda/4$. According to equations~(\ref{elpolsp}) and (\ref{elbisp}), for the same parameters a four times larger gap is obtained in the bipolaron case, $\Delta_\eta^{bp}=\varepsilon_p\;\lambda$. In the small (bi)polaron limit, $d_{bp}\approx a$, the gap in the electron spectrum is independent of $\lambda$, being two times larger for bipolarons ($\Delta_\eta^{bp}=2\Delta_\eta^{pol}=4\varepsilon_p$). For arbitrary $\lambda$, it can easily be verified numerically that the ratio $\Delta_\eta^{bp}/\Delta_\eta^{pol}$ lies between the two limiting behaviors discussed here, $2\leq\Delta_\eta^{bp}/\Delta_\eta^{pol}\leq4$.

\subsection{Adiabatic bipolarons for $U\neq0$\label{Sec001}}

In the adiabatic regime the main contribution to the bipolaron binding energy is described by the adiabatic potential $U_{AD}(\vec u)$, while the kinetic part of the energy, as well as nonadiabatic corrections, contributes much less to the total bipolaron binding energy. For this reason, some important properties of the adiabatic bipolarons may be obtained simply by calculating the equilibrium lattice deformation, corresponding to the minima of $U_{AD}(\vec u)$. The approach that yields the bipolaron behavior from these minima is hereafter referred to as the static adiabatic approximation (SADA), reflecting the fact that only the adiabatic equilibrium point in the lattice configuration space is being considered, while the dynamics are neglected.

\subsubsection{Small bipolarons}

The effects of the Hubbard repulsion on the formation of the small bipolarons have been discussed in numerous works. For $U=0$ and $\lambda$ large, the pair of electrons and the accompanying lattice deformation localize to a single lattice site, forming a so called $S0$ bipolaron \cite{Proville,Bonca,Macridin}. With the weak Hubbard repulsion switched on, the condensation energy is given by $\Delta_{bp}^{S0}\approx2\varepsilon_p-U$. However, for a stronger repulsion $U\gtrsim2\varepsilon_p$ the electrons tend to avoid the same lattice site. The resulting bipolaron consisting of two weakly overlapping small polarons at the nearest neighbor sites is usually referred to as an $S1$ bipolaron \cite{Proville}. By keeping only the leading corrections in $1/U$ and $1/\lambda$ small, the SADA condensation energy of the $S1$ bipolarons takes the form

\begin{equation}
\Delta_{bp}^{S1}\approx4t^2/U-\varepsilon_p/\lambda^2\;.\label{DeltaS1}
\end{equation}

\noindent In fact, $\Delta_{bp}^{S1}$ can easily be interpreted by starting from the $U=\infty$ limit. For $U=\infty$, placing two small polarons next to each other divides the energy gain associated with the adiabatic spreading of the small polaron by half. This repulsive effect is described by the $1/\lambda^2$ correction to the small-bipolaron binding energy $\varepsilon_p$, i.e., by the second term in equation~(\ref{DeltaS1}). Returning to the finite $U$ case, the $S1$ bipolaron is stabilized for $U\lesssim4\varepsilon_p$ by the superexchange energy, given by the first term in equation~(\ref{DeltaS1}).

\begin{figure}[tb]

\begin{center}{\scalebox{0.25}
{\includegraphics{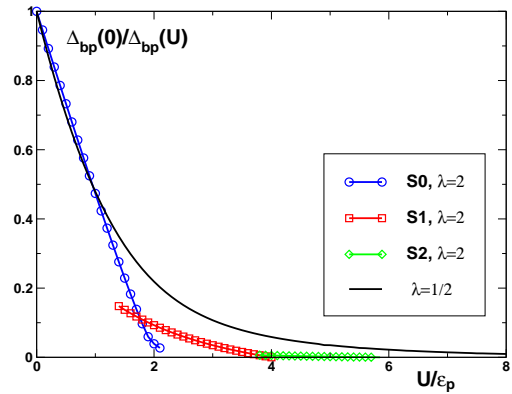}}}
\end{center}

\caption{(Color online) The SADA condensation energy of the small ($\lambda=2$) and large ($\lambda=1/2$) bipolaron as a function of $U/\varepsilon_p$. For the small bipolaron case, different symbols are used for $S0$, $S1$ and $S2$ condensation energies.\label{fig01}}

\end{figure}

In Figure~\ref{fig01} the binding energy of the small bipolaron ($\lambda=2$) is shown as a function of $U/\varepsilon_p$ by curves with symbols. The curves are normalized by the $U=0$ value of $\Delta_{bp}$. For $\lambda=2$, the SADA transition in Figure~\ref{fig01} between the $S0$ (circles) and $S1$ (squares) bipolarons occurs for $U/\varepsilon_p\approx1.8$. The second SADA transition in Figure~\ref{fig01} takes place for $U/\varepsilon_p\approx3.8$, involving a transition between $S1$ (squares) and $S2$ (diamonds) bipolarons. $S2$ denotes a bipolaron consisting of two weakly bound small polarons at next-nearest neighbor sites. Increasing $U$ further, the SADA gives weakly overlapping small polarons at increasing distances ($Si$, $i>2$), with a vanishing binding energy. By calculating the minima of the adiabatic potential as a function of $U$ to the leading correction in $1/\lambda$ small, it may be shown \cite{Dorignac} that two polarons become unbound for a critical value of the Hubbard repulsion $U_c$. In particular, for $1/\lambda=0$ this value is given by $U_c=12\;\varepsilon_p$, with $U_c$ shifting towards larger values as $\lambda$ decreases away from the atomic limit $1/\lambda=0$ \cite{Dorignac}.

Because of strong lattice coarsening effects, the $S0$ and $S1$ bipolarons, corresponding to two minima (local and global) of the adiabatic potential $U_{AD}(\vec u)$, are separated by a substantial energy barrier. Consequently, at the transition between different $Si$ bipolarons ($i=0,1,\ldots$) the SADA condensation energy $\Delta_{bp}$ in Figure~\ref{fig01} exhibits a pronounced singularity in its slope. This behavior should be contrasted to the large bipolaron case $d_{bp}\gg a$ discussed in Section~\ref{Sec001LP}, for which the lattice coarsening effects are negligible and the SADA condensation energy $\Delta_{bp}$ behaves smoothly [see the $\lambda=1/2$ case in Figure~\ref{fig01}].

In general, kinetic and nonadiabatic contributions are expected to smear any singular behavior of the exact condensation energy $\Delta_{bp}$. Indeed, unlike in Figure~\ref{fig01}, $\Delta_{bp}$ calculated numerically by the RCSM in Section~\ref{Sec3} always exhibits a smooth crossover between different $Si$ bipolarons.

\subsubsection{Large bipolarons\label{Sec001LP}}

The problem of the large HH bipolaron has attracted much less attention in the literature than has the problem of the small bipolaron. In the continuum approximation, appropriate for $d_{bp}\gg a$, $U_c=2.5\;\varepsilon_p$ has been reported \cite{Emin,Bussac} as the critical strength of the Hubbard repulsion above which the adiabatic bipolaron is unstable with respect to forming two separate adiabatic polarons. This value has been obtained by a variational technique involving a product of single-electron wave functions.

Here, in the context of the SADA, the condensation energy of the large adiabatic HH bipolaron is studied by calculating the exact minima of $U_{AD}(\vec u)$, without assuming in advance any particular functional form of the adiabatic electron wave function. 

While the behavior of the bipolaron condensation energy $\Delta_{bp}$ in the small bipolaron limit $1/\lambda\rightarrow0$ may be approximately obtained \cite{Dorignac} from the leading corrections in $1/\lambda$ small, the general case encompasses the summation of the whole $1/\lambda$, $1/U$ expansion of the adiabatic binding energy. Within the SADA, this summation is performed numerically, by using an iterative scheme proposed in reference~\cite{Proville}, supplemented with appropriate modifications necessary to preserve numerical stability in the large bipolaron limit. Following this procedure, the stability of large adiabatic bipolarons is established for much larger values than predicted before. 

In Figure~\ref{fig01} the condensation energy $\Delta_{bp}$ of the large adiabatic polaron is shown as a function of $U/\varepsilon_p$ (full curve). Comparing $\Delta(U)/\Delta(0)$ for the small and large bipolaron cases ($\lambda=2$ vs. $\lambda=1/2$), one observes that the condensation energy of the large adiabatic bipolarons drops more slowly. This also means that the critical value of the Hubbard repulsion $U_c$ for the bipolaron stability increases as $\lambda$ decreases, shifting $U_c$ to much larger values ($U_c>12\varepsilon_p$). However, as it may be seen from Figure~\ref{fig01}, irrespectively of $\lambda$, which defines the size of the bipolaron, $\Delta_{bp}$ is rapidly suppressed by $U/\varepsilon_p$, becoming exponentially small for $U/\varepsilon_p\gtrsim4$.

\begin{figure}[tb]

\begin{center}{\scalebox{0.25}
{\includegraphics{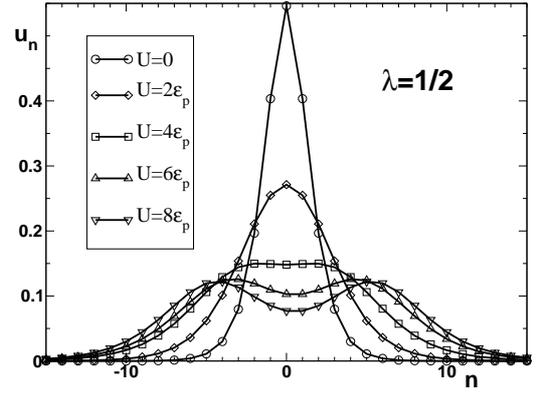}}}
\end{center}

\caption{The SADA lattice deformation of large bipolarons ($\lambda=1/2$) as a function of $U$, showing the dissociation of the bipolaron in the adiabatic limit.\label{fig02}}

\end{figure}

Further insights into the formation of large bipolarons as a function of the Hubbard repulsion $U$ can be obtained by examining the SADA lattice deformation. In Figure~\ref{fig02} this deformation is plotted for $\lambda=1/2$ and various values of $U/2\varepsilon_p=i$, $0\leq i\leq4$. With increasing $U$, the large bipolaron in Figure~\ref{fig02} progressively transforms into a pair of weakly bound large polarons at increasing distance. As may be seen from Figure~\ref{fig02}, the lattice deformation at the central bipolaron site $n=0$ is the largest for $U/\varepsilon_p\lesssim4$. On the other hand, for $U/\varepsilon_p\gtrsim4$ the largest deformation is found for non central sites $n\neq0$. A close inspection shows that, for some values of $U$, the center of the symmetry of the SADA lattice deformation is in fact between two lattice sites, e.g., between the $n=0$ and $n=1$ sites for $U/2\varepsilon_p=3$ and $U/2\varepsilon_p=4$ in Figure~\ref{fig02}. The same symmetry of the lattice deformation is found in the small bipolaron limit for the $S1$ bipolarons.

When $\Delta_{bp}$ obtained by the SADA acquires small values, corrections associated with the lattice kinetic energy are decisive for the stability of the adiabatic bipolaron. The results start to be even more intriguing when nonadiabatic contributions assume an important role. In this context, the interesting case appears when one starts in the $U\rightarrow\infty$ limit with two nonadiabatic polarons. By decreasing $U$ the two polarons condense into a bipolaron, which for $U=0$ may be dominated by the adiabatic dynamics. In order to describe accurately such a kind of crossovers involving a mixture of adiabatic and nonadiabatic dynamics one has to rely on numerical approaches like the RCSM.

\section{Numerical results\label{Sec3}}

In Section~\ref{Sec2} the adiabatic limit was discussed in terms of the SADA results. Beyond this, the application of the RCSM allows a dynamical quantum description of bipolarons as well as the extension of the current study to the whole parameter space. 

It is instructive to start the numerical analysis with a comparison to other methods, when the latter are applicable, so as to establish the accuracy of the RCSM. For $\omega_0=t=g$ the value of the RCSM ground state (zero momentum $K=0$) energy is $E_{bp}=-5.420\;\omega_0$, which is close to the practically exact value of $E_{bp}=-5.424\;\omega_0$ \cite{Bonca}. For the same parameters, high accuracy ($E_{bp}=-5.419\;\omega_0$) is also achieved by a variational method described in reference~\cite{Filippis}.

\subsection{Nonadiabatic contributions}

\begin{figure}[tbh]

\begin{center}{\scalebox{0.3}
{\includegraphics{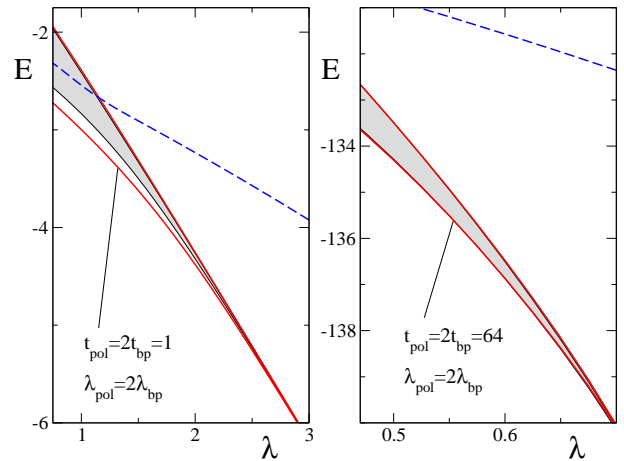}}}
\end{center}

\caption{(Color online) The RCSM lowest band, bounded by $K=0$ and $K=\pi$ states, for the Holstein bipolaron (gray area) and polaron (full thick curves) are compared as a function of $\lambda_{pol}=2\lambda_{bp}$ for $t_{pol}=2t_{bp}=\omega_0$ (left panel) and $64\;\omega_0$ (right panel) fixed (note different $\lambda$ and energy scales in two panels). The dashed curves are twice the RCSM polaron ground state energy for $\lambda=\lambda_{bp}$, defining the energy threshold for the bipolaron stability. ($\omega_0$ is used as the unit of energy.)\label{fig03}}

\end{figure}

The differences between the nonadiabatic dynamics of bipolarons and polarons become evident by comparing the spectra for $U=0$ through the scaling given by equation~(\ref{substi}). For this purpose Figure~\ref{fig03} is used, with RCSM curves obtained by varying $\lambda=\lambda_{bp}=2\lambda_{pol}$, while $t=t_{bp}=t_{pol}/2$ is kept fixed. For the right panel of Figure~\ref{fig03} the ratio $t/\omega_0$ is chosen to be much larger than for the left panel in order to contrast the behavior obtained close to the adiabatic limit with the regime where the nonadiabatic effects play a significant role. For the different choice of $\lambda$ scales, both panels exhibit similar band-narrowing effects. However, due to the very different values of $t/\omega_0$, two different physical mechanisms are involved.

In Figure~\ref{fig03}, the lowest bipolaron band is represented by the gray area, with boundaries defined by the $K=0$ and $K=\pi$ states. This bipolaron band is compared, using the scaling in equation~(\ref{substi}), to the energy of the lowest polaron band, whose boundaries are given by the full thick $K=0$ and $K=\pi$ curves. The dashed curve in Figure~\ref{fig03} is twice the energy of the polaron ground state, plotted for $\lambda=\lambda_{bp}$ and $t=t_{bp}$. It defines the threshold energy for the stability of the bipolaron $K=0$ (ground) state.

For $\lambda_{pol}=2\lambda_{bp}$ and $t_{pol}=2t_{bp}$, in the absence of nonadiabatic contributions, the bipolaron and the polaron bands in Figure~\ref{fig03} should be the same. However, the left panel ($t_{pol}=2t_{bp}=\omega_0$) in Figure~\ref{fig03} clearly shows a larger bandwidth of the polaron band, indicating that the nonadiabatic contributions are more efficient in delocalizing the polarons. As the decrease of the adiabatic gap in the electron spectrum $\Delta_\eta$ leads to an increasing importance of nonadiabatic effects, the differences between the bipolaron and the polaron band become more pronounced towards the weak-coupling (left) side of the left panel in Figure~\ref{fig03} ($\Delta_\eta\lesssim\omega_0$). The explanation is quite simple. In nonadiabatic processes the electron is temporally detached from the lattice deformation. Since two electrons have to work cooperatively in order to nonadiabatically delocalize the bipolaron, it is not surprising that the polaron delocalizes nonadiabatically more efficiently through single electron processes. 

On the contrary, for the large $t/\omega_0$ used in the right panel of Figure~\ref{fig03} ($t_{pol}=2t_{bp}=64\;\omega_0$), the differences between dispersions of the polaron and the bipolaron bands are hardly seen. In other words, the dynamics is almost completely adiabatic. The significant band-narrowing in the right panel of Figure~\ref{fig03} is governed by lattice coarsening effects that become stronger as $\lambda$ increases, rather than by the change in the nature of the electron-phonon correlations (i.e., adiabatic vs. nonadiabatic), as it is in the left panel of Figure~\ref{fig03}.

The qualitative difference between the two (bi)polaron band-narrowing mechanisms in the two panels of Figure~\ref{fig03} may be argued further from the behavior of the bipolaron binding energy $E_{bp}$. This energy in Figure~\ref{fig03} corresponds to the energy difference between the minimal energy of two free electrons $-4t_{bp}$ and the bottom of the bipolaron bands. As one may observe, $E_{bp}$ takes very different values in the left and right panels of Figure~\ref{fig03}. In particular, the small (large) binding energy in the left (right) panel of Figure~\ref{fig03} directly indicates the small (large) gap in the adiabatic electron spectrum $\Delta_\eta\propto E_{bp}$, discussed already in connection with equations~(\ref{elpolsp}) and (\ref{elbisp}). For $\Delta_\eta\lesssim\omega_0$ nonadiabatic dynamics prevails, while $\Delta_\eta\gtrsim\omega_0$ represents the opposite, dominantly adiabatic behavior.

\subsection{Bipolaron band structure}

Depending on parameters, excited coherent bipolaron bands may emerge below the phonon threshold for the incoherent scattering. In fact, due to the energy constraint in situations when the low-frequency coherent bands are sufficiently narrow, coherent bands can also be found above the threshold energy. Namely, with narrow bands at the bottom of the spectrum, some parts of the spectrum above the phonon threshold may remain inaccessible to incoherent phonon excitations that add $\omega_0$ (optical phonon energy) to the total energy of the system. Analogous behaviors of the polaron spectrum have been found in investigations reported in references~\cite{Vidmar,Ciuchi}.

In a manner similar to that discussed for the lowest band in connection with Figure~\ref{fig03}, under the substitution of parameters (\ref{substi}) any differences between the excited polaron and bipolaron bands in the $U=0$ case should be attributed to nonadiabatic effects. Since these differences do not bring any essentially new behavior and since the polaron band structure as a function of the coupling constant $g$ has been extensively reported upon previously \cite{Barisic8,Barisic10,Barisic5}, we turn instead to the role of the Hubbard repulsion.

\subsubsection{Relation to soft normal modes}

By revealing excited bipolaron bands, the RCSM is able to provide a detailed perspective of various aspects of bipolaron formation. The excited bands are associated with the adiabatically softened phonon modes that move along the lattice with the bipolaron. Thus, by analogy with the polarons \cite{Barisic10}, the excited bipolaron bands serve as fingerprints of adiabatic correlations. Generally speaking, the strongest adiabatic correlations should be expected in the low-frequency part of the bipolaron spectrum, for which the corresponding lattice deformation is the slowest. The absence of the excited bipolaron bands below the phonon threshold therefore indicates that, for all frequencies, the dynamics is nonadiabatic. Namely, the electrons detach nonadiabatically from the phonon cloud too frequently and the fluctuations of the lattice at different sites remain adiabatically uncorrelated (non-softened) by electrons. Indeed, in the weak-coupling regime ($\Delta_\eta<\omega_0$) one always finds only the lowest coherent bipolaron band below the phonon threshold. 

\begin{figure}[t]

\begin{center}{\scalebox{0.22}
{\includegraphics{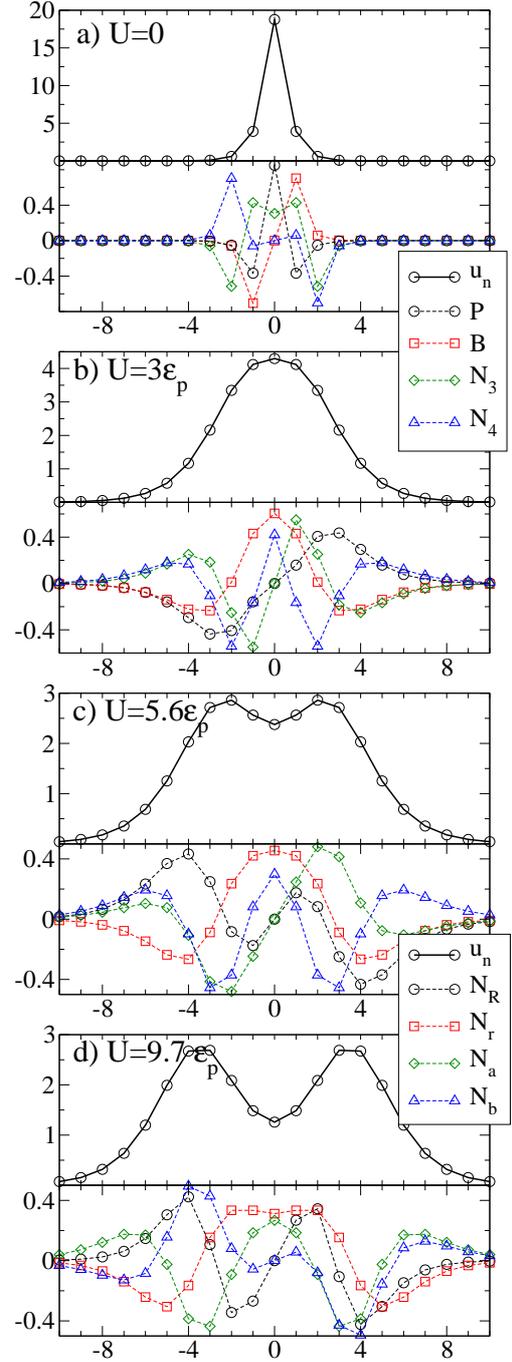}}}
\end{center}

\caption{(Color online) The SADA lattice deformation and the lowest four (soft) normal modes as a function of $U$ ($\lambda=0.98$, $t/\omega_0=200$).\label{fig04}}

\end{figure}

Due to the softening of the phonon modes, the bipolaron band structure starts to be particularly intriguing upon approaching the adiabatic limit. More specifically, by varying $U$, the bipolaron band structure transforms from the polaron-like behavior at $U=0$ [in the sense of equation~(\ref{substi})] to the large $U$ behavior, when the bipolaron consists of two weakly overlapping polarons sharing the same center of mass coordinate. This gradual transformation of the band structure is examined in detail in Figures~\ref{fig04}-\ref{fig06}.

The four panels in Figure~\ref{fig04} show the SADA (equilibrium) lattice deformation $u_n$ and the lowest four adiabatic (soft) normal bipolaron modes. The first two panels in Figure~\ref{fig04} correspond to the entry and exit values of the parameter $U$, as a function of which the bipolaron band structure is plotted in Figure~\ref{fig05}. The two remaining panels in Figure~\ref{fig04} correspond to the entry and exit values of the parameter $U$ for which the band structure is shown Figure~\ref{fig06} and discussed in Section~\ref{LargeU} in the context of the large $U$ limit.

As in Figure~\ref{fig02}, the dissociation of the bipolaron as a function of $U$ can easily be tracked in Figure~\ref{fig04} from $u_n$. The parameter $\lambda$ in Figure~\ref{fig04} is approximately twice as large as in Figure~\ref{fig02} ($\lambda=0.98$ vs. $\lambda=1/2$). Consequently, the bipolarons in Figure~\ref{fig04} exhibit significant lattice coarsening effects, being set by this particular choice of $\lambda$ in the vicinity of the crossover regime between the large and the small adiabatic bipolarons. 

The normal adiabatic modes of the bipolaron lattice deformation field, shown in Figure~\ref{fig04}, are obtained using the harmonic approximation for the adiabatic potential $U_{AD}$ in equation~(\ref{UADU0}). They may be distinguished according to the number of nodes and their parity. Namely, the modes are even or odd under reflection with respect to the center of the equilibrium lattice deformation at $n=0$. Depending on the parameters used, the energies of the modes with different parities may cross. 

Although calculated by breaking the translation symmetry of the adiabatic bipolaron problem, the normal modes in Figure~\ref{fig04} give a clear meaning to the internal structure of the low-frequency adiabatic dynamics. That is, by considering the relationship between the normal mode coordinates and the adiabatic coordinate that corresponds to the translational bipolaron motion along the lattice, one may distinguish between the local dynamics orthogonal to the motion of the center of mass and the translational motion of the center of mass itself. 

\subsubsection{Band structure for small $U$}

\begin{figure}[tb]

\begin{center}{\scalebox{0.31}
{\includegraphics{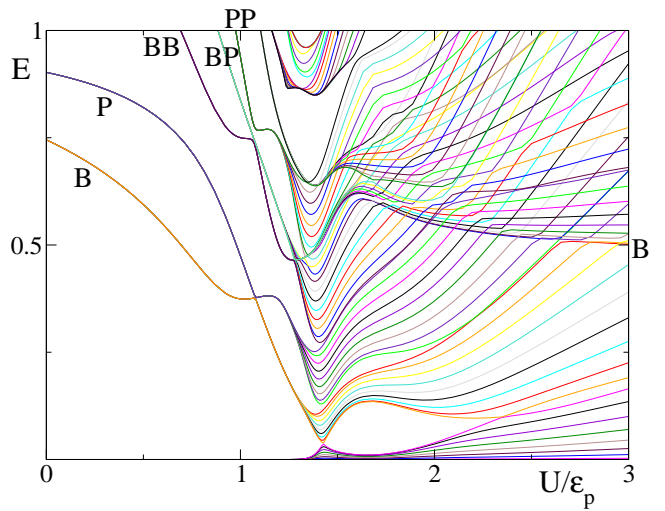}}}
\end{center}

\caption{(Color online) The bipolaron band structure as a function of $U$ ($t/\omega_0=200$, $g=14$, $\lambda=0.98$). The lowest and excited bands corresponding to the excitations of the pinning $P$ and breather $B$ normal modes are shown.\label{fig05}}

\end{figure}

In the small $U$ regime, the main effect of the Hubbard repulsion is to increase the size of the bipolaron. For this reason the bipolaron band structure in Figure~\ref{fig05} bears many resemblances to the small to large polaron crossover \cite{Barisic8,Barisic5} as $\lambda$ is varied. Starting with narrow bands on the left (small bipolaron) side of Figure~\ref{fig05}, as $U$ is increased the widths and the distances between various excited bands become comparable. As $U$ is increased further the band structure on the right side of Figure~\ref{fig05} develops the large bipolaron behavior. 

The bipolaron bands in Figure~\ref{fig05} are given in terms of 9 states with different momenta $K$, $K=m\times\pi/8$, with $0\leq m\leq8$. All the energies are shifted by the ground state energy of the bipolaron. In order to simplify the analysis, additional bands below the phonon threshold associated with higher normal modes are not considered in Figure~\ref{fig05} (e.g., $N3$, $N4$ modes shown in Figures~\ref{fig04}a and \ref{fig04}b, and higher modes).

The detailed understanding of the band structure in Figure~\ref{fig05} involves a distinction between two basic effects \cite{Barisic8,Barisic5}. The first explains the bandwidths, and is related to the so called Peierls-Nabarro (PN) barrier $\Delta_{PN}$. The second explains the hybridization between excited bands and is related to the effective coupling between normal modes.

Regarding the role of the PN barrier, shifting the large adiabatic bipolaron across the unit cell from its exact equilibrium position has a very small energetic cost $\Delta_{PN}\ll\omega_0$. Namely, because the lattice coarsening effects are suppressed for $d_{bp}\gg a$, the shape of the large bipolaron is almost preserved at any point of the minimal energy path for the bipolaron translation that connects the exact minima of the adiabatic potential $U_{AD}(\vec u)$. This effect may be observed in the frequency of the pinning $P$ mode, which vanishes in the $\Delta_{PN}\rightarrow0$ limit. Consequently, the large bipolaron moves along the lattice almost as a free particle. On the other hand, due to the lattice discreteness, the shape of the small bipolaron changes substantially within the unit cell along the minimal energy path for the translation. Therefore, for small bipolarons the PN barrier is large, $\Delta_{PN}\gg\omega_0$, which results in the very narrow bipolaron bands in the spectrum observed in the left part of Figure~\ref{fig05}. Their positions in the spectrum correspond to the excitation energies of the normal modes shown in Figure~\ref{fig04}. At the left side of Figure~\ref{fig05} the bands are denoted accordingly, e.g., the band associated to the simultaneous excitation of the pinning and breather mode is denoted by $BP$. 

The second effect that requires consideration is the effective coupling between the normal modes. The lowest even (breather) mode, denoted by $B$ in Figures~\ref{fig04}a and \ref{fig04}b, involves vibrations of the bipolaron size, whereas the lowest odd (pinning) mode, denoted by $P$, displaces the center of mass. The change in the shape of the small bipolaron along the minimal energy path results in a strong effective coupling between the $P$ and the $B$ mode \cite{Barisic8,Barisic5}. This explains the strong hybridization of the excited bands in the central part of Figure~\ref{fig05}. 

For the large bipolarons in the right part of Figure~\ref{fig05}, gaps between various bands (associated with different kinetic energies of the translational motion) close as $\Delta_{PN}\rightarrow0$, whereas the hybridization between excited bands involving different degrees of freedom (e.g., $P$ and $B$) becomes weak. $B$ on the right side of Figure~\ref{fig05} denotes the $K=0$ state at the bottom of the band associated with the excitation of the breather mode. The position of this $K=0$ state in the spectrum is approximately given (up to the kinematic effects \cite{Turkevich}) by the frequency of the breather $B$ mode of the large bipolaron, shown in Figure~\ref{fig04}b.
 
\subsubsection{Band structure for large $U$\label{LargeU}}

Turning now to large $U$, it is convenient to discuss the bipolaron properties in terms of two overlapping polarons. Starting with Figures~\ref{fig04}, instead of $B$ for breather, $P$ for pinning, etc., it is useful to introduce a new categorization of the normal modes, $N_R$, $N_r$, $N_b$, $N_a$. Here, $N_R$ and $N_r$ can be understood as bonding and antibonding combinations of the two pinning modes corresponding to the two overlapping polarons. The normal mode $N_R$ is odd and displaces the center of the bipolaron mass, whereas the mode $N_r$ is even and describes vibrations of the distance between two polarons. Similarly, the next two normal modes $N_b$ and $N_a$ may be understood as bonding (even) and antibonding (odd) combinations of two breather modes, representing the in phase and antiphase vibrations of the polarons' sizes.

\begin{figure}[tb]

\begin{center}{\scalebox{0.31}
{\includegraphics{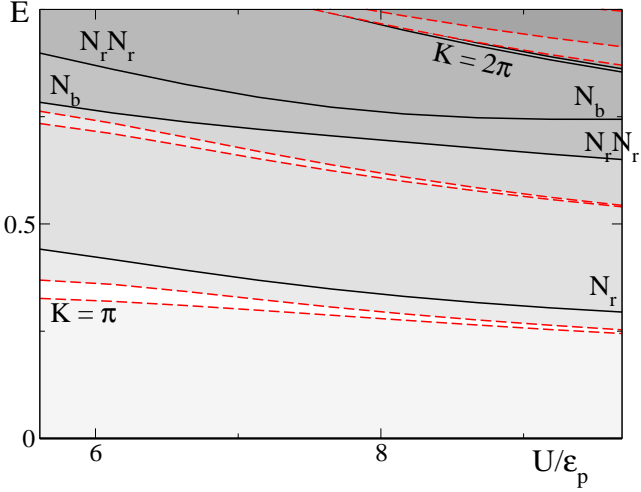}}}
\end{center}

\caption{(Color online) The large bipolaron band structure for large $U$ ($t/\omega_0=200$, $g=14$, $\lambda=0.98$, as in Figure~\ref{fig05}). Beside the bands associated to the translation motion of the bipolaron with zero point motion of the normal modes, the bands with excited normal modes $N_r$ and $N_b$ are shown as well.\label{fig06}}

\end{figure}

The bipolaron band structure for large $U$ is shown in Figure~\ref{fig06}. Steps in gray shading represent the increase in the number of overlapping bands in the spectrum. As in Figure~\ref{fig05}, all energies are shifted by the ground state energy. The band boundaries correspond to the $K=0$ (full curves) and $K=\pi$ (dashed curves) states. Which of the two $K$ states actually defines the bottom and the top of the corresponding band depends on the parity of the states. For example, the ground $K=0$ and the lowest $K=\pi$ state in Figure~\ref{fig05} are even, and define the lower and upper boundaries of the lowest band, respectively. A small gap (white area) separates this lowest band from the first excited band that starts with the $K=\pi$ state of odd parity. In the absence of lattice discreteness effects, the gap between the lowest two bands closes, and the dispersion of the bipolaron states is given simply by $E_{bp}(K)\propto K^2/m_{bp}$, where $m_{bp}$ denotes the bipolaron effective mass. The states associated with the momenta $K=\pi$ and $K=2\pi$ (corresponding to the $N_R$ coordinate) are represented separately in Figure~\ref{fig06}.

Unlike in Figure~\ref{fig05}, where the effective mass $m_{bp}$ of the bipolaron decreases on increasing the bipolaron size with $U$, in Figure~\ref{fig06} $U$ has the opposite effect. Namely, the bandwidth of the lowest band clearly decreases with $U$, indicating that the polaron pair becomes heavier as the mutual distance between the polarons increases.

Beside the bands associated with the increasing kinetic energy of the joint motion of two polarons along the lattice, in Figure~\ref{fig06} additional bands associated to two even normal modes $N_r$ and $N_b$ are shown, with the $K=0$ states denoted according to the nature of the excitation involved. In particular, it may be seen from the $K=0$ state denoted by $N_r$ in Figure~\ref{fig06} that the frequency of the $N_r$ mode decreases with $U$. This behavior is expected on the basis that the restoring force for the vibrations of the distance between two polarons vanishes when the bipolaron dissolves into two unbound polarons. 

It is also worth noting that, by sharing the same parity, the $N_rN_r$ state (double excitation of the $N_r$ mode, $K=0$) and the $N_b$ state (single excitation of the $N_b$ mode, $K=0$), anticross in the central part of Figure~\ref{fig06}. That is, the $K=0$ state denoted by $N_b$ on the left side of Figure~\ref{fig06} changes its nature, being dominantly a double excitation of the $N_r$ mode on the right side of Figure~\ref{fig06}. In other words, the hybridization between bands associated to the $N_b$ and $N_r$ excitations of the bipolaron occurs in Figure~\ref{fig06}. However, in contrast to the hybridization due to lattice coarsening effects in Figure~\ref{fig05}, in Figure~\ref{fig06} one observes only a weak effect of the kinematic \cite{Turkevich} origin, without abrupt changes in the dispersion properties.

\subsection{Light bipolarons with significant condensation energies}

While the adiabatic limit $\Delta_\eta\gg\omega_0$ involves large lattice deformations that make the bipolaron heavy, the weak-coupling limit corresponds to the opposite situation. In this respect, it is interesting to consider which values of $U$ give the most stable, light bipolaron solutions. The existence of light bipolarons with significant condensation energies has attracted particular attention in the context of bipolaron superconductivity \cite{Polaron} and findings that indicate the importance of the electron-phonon interaction in high-T$_c$ materials \cite{expr}. 

In the HH model, one finds \cite{Bonca,Magna,Filippis} that light bipolarons with significant binding energies exist when the relevant energy scales governing the bipolaron dynamics are comparable, $t\sim\Delta_\eta\gtrsim\omega_0$. It is emphasized here that this specific regime of parameters corresponds to the crossover between the weak-coupling limit and the regime of small bipolarons characterized by strong adiabatic correlations and lattice coarsening effects. In other words, light bipolarons with significant binding energies involve a subtle balance of parameters.

\begin{figure}[t]
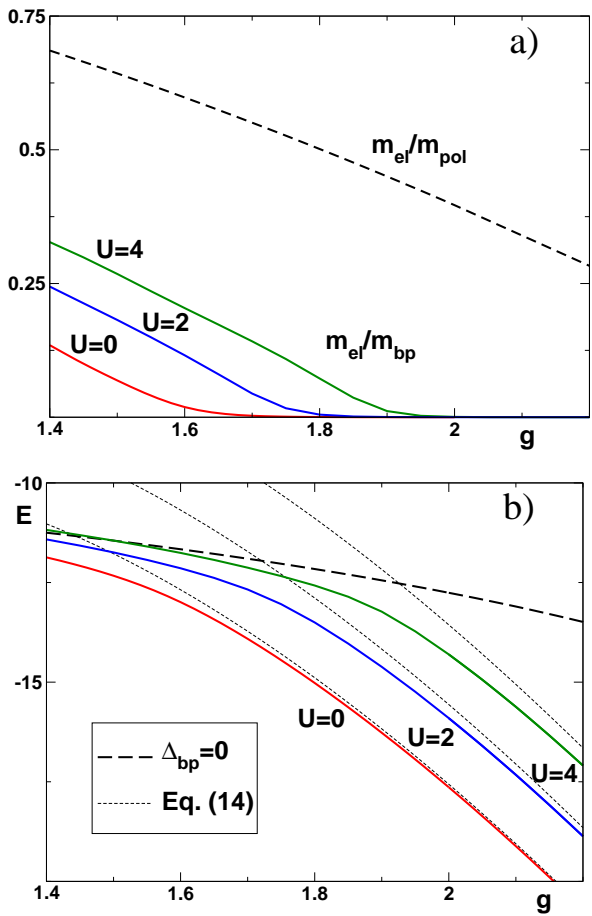


\begin{center}{\scalebox{0.28}{\includegraphics{fig07a.eps}}}
\end{center}
\begin{center}{\scalebox{0.28}{\includegraphics{fig07b.eps}}}
\end{center}

\caption{(Color online) a) The RCSM bipolaron effective mass for $U/\omega_0=0,2,4$. b) The RCSM bipolaron ground-state energy for the same parameters.\label{fig07}}

\end{figure}

Figure~\ref{fig07}a shows the RCSM bipolaron effective mass $m_{bp}$ for three values of the Hubbard repulsion, $U/\omega_0=0,2,4$. The polaron effective mass $m_{pol}$ (dashed curve) is plotted for comparison. A weak renormalization of $m_{pol}$ indicates the weak-coupling regime for the polaron case, i.e., the dynamics are fully nonadiabatic. On the other hand, in contrast to polarons, the bipolaron spectrum (not shown) exhibits excited bands below the phonon threshold for the set of parameters investigated in Figure~\ref{fig07}, meaning that the adiabatic correlations are significant.

The RCSM bipolaron ground-state energy (full curves) is shown in Figure~\ref{fig07}b for the same set of parameters used in Figure~\ref{fig07}a. In the small ($S0$) bipolaron limit ($\varepsilon_p\lambda\gg\varepsilon_p\gg\omega_0$ and $\varepsilon_p\gtrsim U$), corresponding to the right side of Figure~\ref{fig07}b, the ground-state energy approaches asymptotically the values plotted by the thin dashed curves and is given by 

\begin{equation}
E_{bp}\approx -4\varepsilon_p-\varepsilon_p/\lambda^2+U\;.\label{SCPTfit}
\end{equation}

\noindent The first term is the energy of two electrons localized at the same site ($S0$ bipolaron), while the second term is the energy gain due to the adiabatic spreading of the $S0$ bipolaron to the two neighboring sites. Upon substitution of parameters (\ref{substi}), the first two terms in equation~(\ref{SCPTfit}) represent the energy of the small adiabatic polaron calculated to the leading order in small $1/\lambda$. The effects of small Hubbard repulsion in equation~(\ref{SCPTfit}) are taken without $1/\lambda$ corrections, as if two electrons were permanently sharing the same lattice site.

In the crossover towards weak couplings (central part of Figure~\ref{fig07}b), significant deviations from equation~(\ref{SCPTfit}) start to occur. Namely, with decreasing $g$ and/or increasing $U$, the spreading of the lattice deformation renders bipolarons lighter and simultaneously suppresses the gap in the adiabatic electron spectrum $\Delta_\eta$. Such suppression introduces significant nonadiabatic correlations. 

The range of nonadiabatic correlations increases radically in Figure~\ref{fig07}b for states close to the threshold energy for the bipolaron stability, when the bipolaron dissolves nonadiabatically into two polarons. The description of such weak and long-ranged nonadiabatic correlations is quite approximate within the RCSM (and other applicable methods) and small inaccuracies become notable for $\Delta_{bp}\rightarrow0$. In particular, instead of approaching asymptotically the thick dashed $\Delta_{bp}=0$ curve from below in Figure~\ref{fig07}b, the RCSM ground-state energy curve for $U=4\omega_0$ intersects it. Yet, for this particular regime of parameters, the expected relation $m_{bp}\approx2m_{pol}$ is obeyed in Figure~\ref{fig07}a. This result shows that the overall RCSM picture of the bipolaron dissociation for $\Delta_{bp}\rightarrow0$ is semi-quantitatively correct.

\begin{table}[h,t]
\caption{Bipolaron RCSM condensation energy as a function of $U$ for two values of the effective mass. All the energies are in units of $\omega_0$ ($t=2.5\;\omega_0$).\label{table01}}
\begin{center}
\centering{

\begin{tabular}{r|c|c|c}
$m_{bp}/m_{el}=10$&$U=0$&$U=2$&$U=4 $\\\hline\hline 
$g$&$1.44$&$1.58 $&$1.72$\\\hline
$\Delta_{bp}$&$0.71$&$0.38$&
$0.2$\\
\end{tabular}
}\end{center}

\begin{center}
\begin{tabular}{r|c|c|c}
$m_{bp}/m_{el}=20$&$U=0$&$U=2$&$U=4 $\\\hline\hline 
$g$&$1.54$&$1.66$&$1.8$\\\hline
$\Delta_{bp}$&$1.12$&$0.58$&$0.34$\\
\end{tabular}
\end{center}
\end{table}

For the regime $t\sim\Delta_\eta\sim\omega_0$, the relationship between the condensation energy $\Delta_{bp}$ and the effective mass $m_{bp}$ is further elucidated in Table~\ref{table01}. One observes that, for a given effective mass, the condensation energy monotonically decreases with $U$. This means that the most favorable conditions for the formation of very light HH bipolarons with large condensation energies are achieved when the Hubbard repulsion is negligible. 

\section{Entropy considerations\label{Sec1}}

In the HH model the effective interaction between individual bipolarons is repulsive \cite{Bonca}. Therefore, assuming a positive condensation energy $\Delta_{bp}>0$, the electrons tend to condense in pairs. A naive expectation in such circumstances is that the majority of charge carriers condense into bipolarons up to temperatures comparable to $\Delta_{bp}$. However, the entropy of the electron-phonon system depends on the density of charge carriers and the ratio of the concentrations of polarons and bipolarons in the system exhibits a more intricate behavior. Therefore, while the single polaron physics may be invoked for the dilute $\Delta_{bp}<0$ limit, the bipolaron problem $\Delta_{bp}>0$ requires additional considerations. That is, in the dilute limit, even a small temperature $T$ can drive the system from the bipolaronic ground state ($T=0$) to the polaron phase because the latter is favored by the gain in the free energy through the increase of the entropy $S$. This fact is frequently overlooked and, instead, only the binding energy is used to estimate the relative concentrations of polarons and bipolarons.  

In this connection it is instructive to consider the small (bi)polaron limit, for which the length of local electron-phonon correlations $d$ reduces to just one lattice site, $d/a\rightarrow 1$. In this limit, the analysis is considerably simplified because of the vanishing overlaps between bipolarons and polarons, while the Pauli exclusion principle prohibits the double occupancy of lattice sites by electrons with the same spin. If there are $N$ lattice sites, the number of permutations of placing two electrons involved in small polarons on the lattice is $N(N-1)/2$, whereas the small bipolaron can be placed at $N$ different sites. It follows that the gain in entropy of forming two polarons increases logarithmically with the system size $S\sim\ln(N)$. The excited states of the bipolaron do not play a fundamental role here, since the number of them is finite, independent of $N$ and given approximately by $ d_{bp}\Delta_{bp}/a \omega_0$.

For a given total charge concentration $c$ per site the average concentration per site of small bipolarons $c_{bp}$ and the spin-degenerate concentrations of small polarons $c_{\uparrow\downarrow}$ which satisfy $c=2c_{bp}+c_{\uparrow}+c_{\downarrow}$, can be expressed by \cite{Emin002}

\begin{equation}
2c_\uparrow(1-e^{\beta\Delta_{bp}})=1-(1-c)\sqrt{1+\frac{c(2-c)}{(1-c)^2} \;e^{\beta\Delta_{bp}}}\;,\label{cbp}
\end{equation}

\noindent with $\beta$ being the inverse temperature. Although derived previously \cite{Emin002}, some important aspects of equation~(\ref{cbp}) remain to be emphasized. In particular, for any finite $\beta$ in the dilute limit $c\rightarrow0$ of equation~(\ref{cbp}), all the charge is assigned to the polarons \cite{Foot01}, 

\begin{equation}
2c_\uparrow\approx\frac{1-(1-c)(1+c\;e^{\beta\Delta_{bp}})}{(1-e^{\beta\Delta_{bp}})}\approx c\;.
\end{equation}

\noindent With the increase of the total charge concentration $c$, the ratio $c_{\uparrow}/c$ rapidly decreases provided that $\beta\Delta_{bp}$ is large. For example, for $\beta\Delta_{bp}=10$ and $c=0.05$ the ratio $c_{bp}/c$ is close to $0.97$. Yet, for smaller condensation energies like $\beta\Delta_{bp}=5$, large relative values of bipolaron concentration $c_{bp}/c>0.8$ are obtained for $c>0.2$. Thus, the observation of bipolarons in the dilute limit is possible only for low enough temperatures $\beta\Delta_{bp}\gg1$. Otherwise, only polarons will be observed.

In general, for $\Delta_{bp}>0$ and $d_{bp}$ arbitrary, overlapping polarons and bipolarons are simultaneously present in the system, which complicates the estimation of their ratio as a function of doping and temperature. Nevertheless, the free energy gain related to the formation of two polarons instead of the bipolaron is large whenever the correlation length satisfies $d_{bp}/a\ll c^{-1}$.

\section{Summary\label{SecSummary}}

The current work provides a thorough examination of the low-frequency properties that characterize the formation of bipolarons within the 1D Holstein-Hubbard model. Particular attention is devoted to the analogies and differences with respect to polarons, since the properties of these two kinds of quasiparticles are expected to govern the behavior of the electron-phonon system in the dilute limit.

For both, the bipolarons and the polarons, the deep dichotomy in the electron-phonon correlations is fundamentally the same, i.e., adiabatic vs. nonadiabatic. The first interesting observation along these lines is that for $U=0$ the adiabatic bipolarons and polarons exhibit the same spectrum under the simple scaling of parameters, derived here in equation~(\ref{substi}). It is next argued that, for a given set of parameters, bipolarons are always more adiabatic than polarons. This allows some easy predictions of the bipolaron behavior using the already known polaron behavior. 

As a function of $U$, two basic limiting behaviors may be distinguished. For $U$ small, the repulsion between electrons increases the bipolaron size and, depending of parameters, one may observe a small to large bipolaron crossover that is very similar to the small to large polaron crossover when the polaron size is varied through $\lambda=g^2/t\;\omega_0$. For $U$ large, the bipolaron may be discussed in the picture of two overlapping polarons that move together along the lattice. As shown here, detailed aspects of the small and large $U$ behavior may be easily understood from the bipolaron band structure. The latter, including the previously unreported excited bands, is calculated by the RCSM, a method that has previously been successfully applied to the polaron problem. As for polarons, the excited bipolaron bands are associated with the adiabatically softened phonon modes of the moving lattice field. When the weak-coupling regime is achieved, the softening effects are suppressed completely by the nonadiabatic dynamics and the bipolaron spectrum below the phonon threshold involves only the lowest band, just as in the polaron case.

For weak electron-phonon couplings the bare interaction between two electrons can be approximated by an instantaneous effective interaction (the frequency dependence of the phonon propagator may be neglected). Consequently, it may be determined that the electron pair binds for $U\lesssim2\varepsilon_p$. In the adiabatic limit, due to retardation effects associated with the lattice, the bipolaron condensation energy $\Delta_{bp}$ remains positive for large values of $U$. However, for $U\gtrsim4\varepsilon_p$, $\Delta_{bp}$ stays small in the physically relevant part of the parameter space. This property is independent of the bipolaron size, characterizing both the small and the large adiabatic bipolarons.

Particularly interesting is the behavior of small light bipolarons, whose condensation energy is comparable to the bare phonon energy $\omega_0$. Under these conditions, one finds that a fine balance is achieved between the adiabatic, nonadiabatic, and lattice coarsening effects. The role of the Hubbard repulsion is to decrease the effective mass $m_{bp}$ and to suppress $\Delta_{bp}$. In particular, for fixed $m_{bp}$, $\Delta_{bp}$ decreases monotonically with $U$. This implies that the light bipolarons are most strongly bound when the Hubbard repulsion is negligible.

Within the HH model, the effective repulsion between individual bipolarons suppresses the phase separation at low charge densities. Yet, the conditions under which a strong fraction of bipolarons can be observed are severely restricted by the temperature and total charge concentration. For low temperatures $\beta\Delta_{bp}\ll1$, the relative concentration of bipolarons and polarons changes in favor of the former with increasing total charge concentration. On the other hand, for a given finite temperature and a vanishing total concentration, the system is driven into a polaronic phase due to the loss in entropy associated with the formation of bipolarons.

\begin{acknowledgement}

This work was supported by the Croatian Government under Projects $035-0000000-3187$ and $119-1191458-0512$.
 
\end{acknowledgement}


\begin{thebibliography}{99}
\bibitem{Foot01} By considering the dissociation of a bipolaron into two polarons as a kind of chemical reaction, the same limiting behavior would be obtained from the law of mass action.
\end{thebibliography}
\end{document}